\documentstyle[epsfig,amstex,11pt]{article}
\newtheorem{teo}{Theorem}[section]
\newtheorem{lem}{Lemma}[section]

\newtheorem{fed}{Definition}[section]
\newtheorem{con}{Conjecture}[section]

\newcommand{\Eq}[1]{(\ref{#1})}
\newcommand{\Th}[1]{[\ref{#1}]}

\newcommand{\proof} {Proof:}
\def \gl{\mbox{GL}}
\def \R{{\Bbb R}}
\def \C{{\Bbb C}}
\def \Z{{\Bbb Z}}

\def \qed{\hbox{\vrule height 8pt width 8pt}}

\begin{document}
\title{Quadratic Volume Preserving Maps}
\date{May 16, 1997}
\author{H\'ector E. Lomel\'\i\\ James D. Meiss%
\thanks{Useful conversations with R. Easton, K. Lenz and B. Peckham are
gratefully aknowledge. JDM was supported in part by NSF grant
number DMS-9623216.}\\
Department of Applied Mathematics\\
University of Colorado\\
Boulder, CO 80309}

\maketitle
\begin{abstract} We study quadratic, volume preserving diffeomorphisms
whose inverse is also quadratic.  Such maps generalize  the H\'enon
area preserving map and the family of symplectic quadratic maps studied by
Moser.  In particular, we investigate a family of quadratic volume
preserving  maps in three space for which  we find a normal form and
study invariant sets. We also give an alternative proof of a 
theorem by Moser classifying quadratic symplectic maps.
\subsection*{AMS classification scheme numbers:}
34C20,34C35,34C37,58F05,70H99
\end{abstract}

\section{Introduction}

The study of the dynamics of polynomial mappings has a long history
both in applied and pure dynamics.  For example, such mappings provide
simple models of the motion of charged particles through magnetic
lenses and are often used in the study of particle accelerators
\cite{Dragt}.  Moreover, the quadratic, area preserving map,
introduced by H\'enon \cite{Henon69}, is one of the simplest models of
chaotic dynamics.

H\'enon's study can be generalized in several directions. For
example, Moser \cite{Moser} studied the class of quadratic, symplectic
maps, obtaining a useful decomposition and normal form. Here
we do the same for more general class of quadratic, volume preserving
maps, with one caveat as we discuss below.

Just as symplectic maps arise as Poincar\'e maps of Hamiltonian flows,
volume preserving maps are obtained from incompressible flows, and as
such have application to fluid and magnetic field line dynamics
\cite{Holmes, Lau}. Moreover, one can argue that computational
algorithms for flows should obey the ``principle of qualitative
consistency'' \cite{Thy}: if a flow has some qualitative property then
the algorithm should as well.  For the case of Hamiltonian flows this
leads to the construction of symplectic algorithms.  A volume
preserving algorithm should be used for a volume preserving flow, such
as the motion of passive particle in an incompressible fluid
\cite{Kang, Quispel, Shang, Suris}.

Some of the properties of symplectic maps generalize to the
volume preserving case.  For example, a map that is sufficiently close
to integrable and nondegenerate in a certain sense has lots of
codimension one invariant tori \cite{Sun,Cheng,Xia92}, which are
absolute barriers to transport \cite{Fein}.  Also, a perturbation of a
volume preserving map with a heteroclinic connection can have an
exponentially small transversal crossing \cite{Rom-Kedar}.  Finally,
the Birkhoff normal form analysis can be used to study the motion in
the neighborhood of fixed points \cite{Bazzani}

Another motivation for the study of volume preserving maps is that
they can be used as simple models for the study of transport in higher
dimensions.  The general theory of transport is based on a partition
of phase space into regions between which transport is restricted by
partial barriers \cite{MMP84}.  For example, in two dimensions a
partition is formed from intersecting stable and unstable manifolds of
a saddle periodic orbit.  In higher dimensions an analogous
construction requires the existence of codimension one manifolds that
separate the space \cite{MacKay}.  In most cases it is difficult to
find a dynamically natural construction of such manifolds; however,
such manifolds do appear in volume preserving maps, and this leads
easily to the construction of partial barriers.

The computation and effective visualization of invariant manifolds in
higher dimensional maps is itself an interesting problem \cite{Gillian}.
In this paper we will study the intersections of the two dimensional
stable and unstable manifolds in $\R^{3}$.

Polynomial maps are also of interest from a mathematical perspective.
Much work has been done on the ``Cremona maps,'' that is polynomial
maps with constant Jacobians \cite{Engel58}.  An interesting
mathematical problem concerning such maps is the conjecture proposed
by O.T.  Keller in 1939:

\begin{con}[Real Jacobian Conjecture]
 Let $f: \R^n \rightarrow \R^n$ be a
Cremona map. Then $f$ is bijective and has a polynomial inverse.
\end{con}

This conjecture is still open.  It is known that injective polynomial
maps are automatically surjective and have polynomial inverses
\cite{Rudin, Bass}, so it would suffice to prove that $f$ is injective.  It
is easy to see (cf.  lemma \Th{cui} below) that for the quadratic
case, the condition of volume preservation implies injectivity, thus
the Jacobian conjecture holds for quadratic maps.

Even if the conjecture is true, the degree of the inverse of a Cremona
map could be large.  For example, the upper bound for the degree of
the inverse of a quadratic map on $\R^n$ is known to be $2^{n-1}$
\cite{Bass}.  Thus in two dimensions the inverse of a quadratic area
preserving mapping is quadratic, as was discussed by H\'enon
\cite{Henon76, Engel58}.  More generally, Moser showed that quadratic
symplectic mappings in any dimension have quadratic inverses
\cite{Moser}.

H\'enon found the normal form for the quadratic Cremona mapping in the
plane.  In this paper, we will correspondingly find the normal form
for higher dimensional cases, but we assume that the quadratic,
volume preserving mapping has a quadratic inverse (it is a ``quadratic
automorphism'').  We give a complete classification of these
diffeomorphisms.

Such maps can be written as the composition of an affine volume
preserving map and a ``quadratic shear.'' We give necessary and
sufficient conditions for such shears to have a quadratic inverse.  As
a first application of this concept, we give a simple proof of the
theorem of Moser \cite{Moser} for the symplectic case.

We also show that the quadratic automorphism in $\R^3$ can be reduced
to one of three normal forms.  The generic case has four parameters:
two govern the linearization of the map--the trace and second trace of
the Jacobian matrix.  The remaining two parameters determine the
nonlinear terms represented by a single quadratic form in two
variables.
\section{Quadratic Shears}
In this section we will study maps of the form \[x\mapsto x+\frac12
Q(x)\] where $Q$ is a vector of quadratic polynomials.  Throughout
this paper we will write vectors of quadratic polynomials using the
form $Q(x)=M(x)x$ where $M:\R^n \to \gl(\R^n)$ is a linear function
into the set of $n\times n$ matrices that satisfies the symmetry
property $M(x)y=M(y)x$ so that $D_x(M(x)x)=2M(x)$.

\begin{fed}
We say that $f:\R^n\to\R^n$ is a quadratic map in standard form
if $f$ can be written as
\[
     f(x)=x+\frac12M(x)x
\]
where $M$ is a matrix valued linear function that satisfies
$M(x)y=M(y)x$.
\end{fed}
It is important to notice that $Df(x)=I+M(x)$.


\begin{lem} \label{cui}
Let $f(x)=x+\frac12M(x)x$ be a quadratic map of $\R^n$
in standard form. The following statements are equivalent
\begin{enumerate}
\item For all $x\in\R^n$, $\det(Df(x))=1$.
\item $f$ is bijective with polynomial inverse.
\item $\left[M(x)\right]^n=0$.
\end{enumerate}
\end{lem}

\proof

We will show iii)$\Rightarrow$ii)$\Rightarrow$i)$\Rightarrow$iii).
\subparagraph*{iii)$\Rightarrow$ii)}
The condition implies that the matrix $I+M(x)$ is invertible with
inverse $I-M(x)+M(x)^2-\cdots-(-1)^nM(x)^{n-1}$.
We notice that we can write
\[
   f(x)-f(y)=\left(I+M\left(\frac{x+y}2\right)\right)(x-y).
\]
So the function is injective. Using theorem A in \cite{Rudin}, we conclude
that $f$ is bijective with a polynomial inverse.

\subparagraph*{ii)$\Rightarrow$i)}
In principle, $\det(Df(x))$ and  $\det(Df^{-1}(f(x)))$ are
 polynomials in $x_1,x_2,\ldots,x_n$. However, differentiation of
 $f^{-1}(f(x))=x$ gives
\[
   \det(Df^{-1}(f(x)))\det(Df(x))=1,
\]
 and therefore, since both are polynomials,
 $\det(Df(x))$ has to be a constant independent of
$x$. We notice that $\det(Df(x))=\det(Df(0))=\det(I)=1$.

\subparagraph*{i)$\Rightarrow$iii)}
Since $\det(I+M(x))=1$ and $M$ is linear in $x$, then for any $\zeta\neq0$
\[
 \det(M(x)-\zeta I)=(-1)^n\zeta^n\det(I+M(-\frac{1}{\zeta}x))=(-1)^n\zeta^n.
\]
This implies that the characteristic polynomial of $M(x)$ is $(-\zeta)^n$
and therefore $\left[M(x)\right]^n=0$.
\qed

At this point, we restrict to the case of quadratic maps in standard
form whose inverse is also quadratic.  We will see that the dynamics
of such maps is essentially integrable, and is similar to the dynamics
of a shear.  We first establish the following characterization.

\begin{lem}\label{esen}
Let $f(x)=x+\frac12M(x)x$ be a bijective quadratic map of $\R^n$. Then the
following statements are equivalent.
\begin{enumerate}
\item $f^{-1}$ is a quadratic map.
\item $M(x)^2x\equiv0$.
\item $M(x)M(y)z+M(y)M(z)x+M(z)M(x)y\equiv0$.
\item $\displaystyle f^k(x)=x+\frac{k}2 M(x)x$ for all $k\in {\Z}$.
\end{enumerate}
\end{lem}

\proof

We will show i)$\Rightarrow$ii)$\Rightarrow$iii)$\Rightarrow$iv)$\Rightarrow$i).
\subparagraph*{i)$\Rightarrow$ii)} Let \[f^{-1}(x)=x+\frac12
N(x)x\]
where $N(x)$ is a matrix valued linear function that satisfies
$N(x)y=N(y)x$. Then we have that
\[x=f(f^{-1}(x))=x+\frac12 N(x)x+\frac12 M(x+\frac12 N(x)x)(x+\frac12 N(x)x)\]
\[=x+\frac12 N(x)x+\frac12 M(x)x+\frac12 M(x)N(x)x+\frac18M(N(x)x)N(x)x\]
This implies that $N(x)x=-M(x)x$ and $M(x)M(x)x=0$.

\subparagraph*{ii)$\Rightarrow$iii)} By linearity and symmetry of
$M$.

\subparagraph*{iii)$\Rightarrow$iv)} Let
\(\displaystyle g_k(x)=x+\frac{k}2 M(x)x.\)
Then
\[g_k(g_l(x))=x+\frac{l}2 M(x)x+
\frac{k}2 M(x+\frac{l}2 M(x)x)(x+\frac{l}2 M(x)x)\]
\[=x+\frac{l+k}2 M(x)x+\frac{kl}2M(x)M(x)x+
\frac{kl^2}8M(M(x))M(x)x\]
\[=x+\frac{l+k}2 M(x)x-\frac{kl^2}8\left[
	M(x)M(x)M(x)x+M(x)M(M(x))x\right]\]
\[=x+\frac{l+k}2 M(x)x.\]
Therefore $g_k\circ g_l=g_{k+l}$. On the other hand
$g_1=f$ and $g_0=id$. This implies that $g_k=f^k$.
\subparagraph*{iv)$\Rightarrow$i)} Clear.
\qed

\begin{fed}
Let $f:\R^n\to\R^n$ be given, in standard form,
 by $f(x)=x+\frac12M(x)x$.
If $f$ satisfies any of the conditions of lemma \Th{esen}, we
will say that $f$ is a quadratic shear.
\end{fed}

A simple family of quadratic shears is determined by any vector
$v\in\R^n$ and a symmetric matrix $P$ such that $Pv=0$ according to
$M(x)y=(x^TPy)v$, for then
\[
    M(x)^2x=(x^TPv)(x^TPx)v=0.
\]
We will see that, at least in the case $n=3$, this is the most general
quadratic shear.  Moser's normal form for symplectic, quadratic maps
\cite{Moser} shows that the higher dimensional case is not quite this
simple.  From now on, we will concentrate on the special case $n=3$.

\begin{teo} \label{main}
A function $f:\R^3\to\R^3$ is a quadratic shear in $\R^3$ if and only
if there is a vector $v\in\R^3$ and a $3\times3$ symmetric matrix $P$
such that $Pv=0$ and
\[
  f(x)=x+\frac12(x^TPx)v
\]
\end{teo}

\proof

Since $f$ is a bijection, we can define a new function $g:S^2\to S^2$.
on the unit two dimensional sphere $S^2\subset\R^3$, in the following way.
\[g(x)=\frac{f(x)}{|f(x)|}.\]
Using standard theorems of algebraic topology \cite{Green},
 we can argue that
$g$ has either a fixed point or an antipodal point (a point such that
$g(x)=-x$).
In any case,  there is a constant
$K\in{\R}\setminus\{0\}$ and a
vector $x_0\neq0$ such that $f(x_0)=Kx_0$. We will show that
$K=1$. We notice that $x_0$ satisfies the following
\[f(x_0)=Kx_0=x_0+\frac12M(x_0)x_0,\]
\[f^{-1}(Kx_0)=x_0=Kx_0-\frac{K^2}2M(x_0)x_0,\]
and therefore
\[K^2(Kx_0-x_0)=\frac{K^2}2M(x_0)x_0=Kx_0-x_0.\]
We conclude that $K$ satisfies $K^3-K^2-K+1=0$ and hence
$K=1$ or $K=-1$. Clearly, since
$f$ is a bijection, $f(\frac12x_0)\neq0$  implies that
$M(x_0)x_0\neq-4x_0$. We conclude that $f(x_0)\neq-x_0$ and therefore $K=1$.

It is clear that $M(x_0)x_0=0$. Without loss of generality, we can assume that
$x_0=e_1=(1,0,0)$. Notice that $M(e_1)$ has to have the form
\[M(e_1)=\left(\begin{array}{ccc}0&\gamma_1&\gamma_2\end{array}\right)\]
where $\gamma_1,\gamma_2\in\R^3$.
This fact, together with lemma \Th{esen},  implies that the matrix
$M(e_1)^2=0$, and therefore $\gamma_1$ and $\gamma_2$ are parallel. We can
perform a linear change of coordinates and assume without loss of generality
that $f$ has the form
\[
f\left(\begin{array}{c}x_1\\ x_2\\ x_3\end{array}\right)=
\left(\begin{array}{c}x_1\\ x_2\\ x_3\end{array}\right)
+x_1x_2\left(\begin{array}{c}\alpha\\0\\\beta\end{array}\right)+
\]
\[
\frac{x_2^2}2\left(\begin{array}{c}\mu_1\\\mu_2\\ \mu_3\end{array}\right)+
x_2x_3\left(\begin{array}{c}\nu_1\\ \nu_2\\ \nu_3\end{array}\right)+
\frac{x_3^2}2\left(\begin{array}{c}\eta_1 \\ \eta_2\\ \eta_3\end{array}\right)
\]
Let $M_1=M(e_1), M_2=M(e_2)$ and $M_3=M(e_3)$ where
$e_1=(1,0,0), e_2=(0,1,0)$ and $e_3=(0,0,1)$. It is easy to see that
\[
M_1=\left(\begin{array}{ccc}
0&\alpha&0\\
0&0&0\\
0&\beta&0
\end{array}\right)
\]
\[
M_2=\left(\begin{array}{ccc}
\alpha&\mu_1&\nu_1\\
0&\mu_2&\nu_2\\
\beta&\mu_3&\nu_3
\end{array}\right)
\] and
\[
M_3=\left(\begin{array}{ccc}
0&\nu_1&\eta_1\\
0&\nu_2&\eta_2\\
0&\nu_3&\eta_3
\end{array}\right).
\]

To finish the proof, we need to show that the column vectors $\mu,\nu,\eta$ and
$(\alpha,0,\beta)$ of $M_1,M_2,M_3$ are parallel to each other.
We will show step by step
that several of the entries are zero. We have two cases.

\begin{itemize}
\item $\beta\neq0$.
Using lemma \Th{esen} we conclude that $2M_3^2e_1+M_1M_3e_3=0$. This implies
that $\eta_2=0$. We also have that $M_3^3=0$, so $\nu_2=0$ and
$\eta_3=0$. The condition $M_2^2e_2=0$ implies that $\mu_2=0$ and
this together with $M_2^3=0$ implies that $\nu_3=-\alpha$. Using the
equation $M_2M_3e_3+2M_3^2e_2=0$ we find that $\eta_1=0$.
Using $M_2^3=0$ and $M_2^2e_2=0$, we find that the column vectors of
$M_2$ are parallel, and the rest is clear.

\item $\beta=0$.
The condition $M_2^3=0$ implies that  $\alpha=0$,
$\nu_3=-\mu_2$ and $\mu_2^2+\nu_2\mu_3=0$.
 The condition $M_3^3=0$ implies that $\eta_3=-\nu_2$ and
 $\nu_2^2-\mu_2\eta_2=0$. On the other hand,
 $M_2^2e_2=0$ implies that $\mu_1\mu_2+\mu_3\nu_1=0$ and
 $M_3^2e_3=0$ implies that $\nu_1\eta_2-\eta_1\nu_2=0$.

 So it is enough to show that $\mu_1\nu_2-\nu_1\mu_2=0$ and
 $\nu_1\nu_2-\eta_1\mu_2=0$. Clearly, if $\eta_2=0$ then
 $\mu_2=0$ and we would be done. So, we can assume that
 $\eta_2\neq0$.

 If $\eta_2\neq0$ then $\eta_1\mu_2=\eta_1\nu_2^2/\eta_2=\nu_1\nu_2$.
 If $\nu_2=0$ then  $\mu_2=0$ and we would be done. Assume that
 $\nu_2\neq0$ and $\eta_2\neq0$. This implies that
 $\mu_1\nu_2=\mu_1\mu_2\eta_2/\nu_2=-\mu_3\nu_1\eta_2/\nu_2=
 \mu_2^2\nu_1\eta_2/\nu_2^2=\mu_2\nu_1.$
\qed

\end{itemize}


\section{Quadratic Symplectic Maps}

In this section we use the characterization of quadratic shears in 
lemma \Th{esen} to give an alternate proof of the result of Moser
\cite{Moser} for quadratic symplectic maps.  Recall that a map $f$ is
symplectic if $\omega(Dfv,Dfv') = \omega(v,v')$ for all vectors $v,v'
\in \R^{2n}$ where $\omega$ is the standard symplectic form
$\omega(v,v')=v^T J v'$ and $J$ is the $2n\times 2n$ matrix,
\[
	J=\left(\begin{array}{cc}
		0&I\\
		-I&0\\
	\end{array}\right)  \ .
\]
\begin{teo}
Let $f$ be a quadratic symplectic map of $(\R^{2n},\omega)$.  Then $f$
can be decomposed as $f=T\circ S$ where $T$ is affine symplectic and
$S$ is a symplectic quadratic shear.  Furthermore, if $S$ is any
symplectic quadratic shear, then there is a symplectic linear map
$\lambda$ such that $\lambda\circ S\circ\lambda^{-1}(q,p)=(q+\nabla
V(p),p)$.
\end{teo}

\proof

Let $b=f(0)$ and $L=Df(0)$.  Clearly $L$ is a symplectic matrix and if
we let $T(x)=Lx+b$ then $S=T^{-1}\circ f$ is a symplectic quadratic
map in standard form, i.e. $S(x)=x+\frac12M(x)x$, where $M(x)$ is
linear in $x$ and satisfies the symmetry property $M(x)y=M(y)x$.  Then
$S$ is symplectic providing
\[
	(I+M(x))^TJ(I+M(x))=J \ .
\]
Homogeneity of $M(x)$ implies that
\begin{equation}\label{amad}
	M(x)^TJ=J^T M(x) \ ,
\end{equation}
and
\begin{equation} \label {bmad}
	M(x)^TJM(x)=0  \ .
\end{equation}
Using \Eq{amad} in \Eq{bmad} gives $0 = M(x)^{T}JM(x)=J^{T}M(x)M(x)$,
and since $J$ is nonsingular this implies
\begin{equation}\label {cmad}
   M(x)^{2} = 0 \ .
\end{equation}
Then lemma \Th{esen} implies that $S$ is a quadratic shear.

To finish the proof, we follow Moser \cite{Moser} and define the null
space of $M(x)$ in the following way
\[
	N=N(M)=\{y\in\R^{2n}:M(x)y=0, \forall x\in\R^{2n}\} =
	\{y\in\R^{2n}:M(y)=0\}\ .
\]
Recall \cite{Abraham} that the $\omega-$orthogonal complement of a
subspace $E\subset\R^{2n}$ is defined by
$E^\perp=\{v\in\R^{2n}:\omega(v,v')=0,\forall v'\in E \}$.  We will
show that $N^\perp\subset N$. For that purpose, we will use the
following fact 
\begin{equation} \label{ggg}
M(z)M(x)y= M(x-y)^2z=0
\end{equation}
that follows from lemma \Th{esen}, linearity  and symmetry.

Let $u\in N^\perp$ and $x\in\R^{2n}$. 
Now for any $y\in \R^{2n}$, \Eq{ggg}
implies that $M(x)y\in N$.  Therefore 
$\omega(y,M(x)u)=y^TJM(x)u=-y^TM(x)^TJu=-\omega(M(x)y,u)=0$.  This
implies that $M(x)u=0$ and hence $u\in N$.  Standard theorems in
symplectic geometry (cf.  \cite{Abraham}) imply that it is possible to
find a Lagrangian space $F$ such that $N^\perp\subset F^\perp=F\subset
N$ and a symplectic linear transformation $\lambda$ such that
\[
	\lambda (F)=\{(q,p)\in\R^{n}\times\R^{n}:p=0\}.
\]
Clearly, if $S(x)= I+\frac12 M(x)x$ is a symplectic quadratic shear, then
so is $\tilde S=\lambda\circ S\circ\lambda^{-1}$.  Assume that $\tilde
S(x) = I +\frac12 \tilde M(x)x$.  Then $\lambda (F)\subset N(\tilde M)$.
Therefore for all $(q,p)\in\R^{n}\times\R^{n}$,
\[
	\tilde M(q,p)(q,p)=\tilde M(q,p)(0,p)=\tilde M(0,p)(q,p)
	                  =\tilde M(0,p)(0,p) \ .
\]
Since, in general, the matrix $\tilde M(0,p)$ can be written 
in $n \times n$ blocks as
\[
	\tilde M(0,p)=\left(\begin{array}{cc}
						A(p)&B(p)\\
						C(p)&D(p)\\
			   	  \end{array}\right) \ ,
\]
then $\tilde M(0,p)(q,0) = 0$ implies $A(p)=C(p) = 0$. Moreover, \Eq{amad}
implies $D(p)=0$ and $B(p)^T=B(p)$. Thus finally we see that
\[
	 \tilde M(q,p)(q,p) = (B(p)p,0)
\]
where $B(p)p$ is a gradient vector field.
\qed
\section{Normal Form in $\R^{3}$}

In this section we give normal forms for a quadratic diffeomorphism
$f$ of $\R^3$ that preserves volume and has a quadratic inverse. Now
lemma \Th{esen} implies that if we let $b = f(0)$ and $L = Df(0)$, and
$T(x) = Lx+b$, then the map $S = T^{-1}\circ f$ is a quadratic shear. Then
theorem \Th{main} implies that $S$ is of the form
$S(x)=x+\frac12(x^TPx)v$ where $v\in\R^3$ and $P$ is a symmetric
matrix such that $Pv=0$.  Depending on the relation between $L$ and
$v$, we have three cases possible cases; these can by distinguished by
considering the space
\[
	Z(v,L)=span\{v,Lv,L^2v\}  \ .
\]

\begin{teo}\label{normalform}
Let $f:\R^3\to\R^3$ be a quadratic volume preserving diffeomorphism.
Then $f$ can be written as the composition of an affine map $T$ and a
quadratic shear $S$, $f= T \circ S$, where $S(x)=x+\frac12(x^TPx)v$,
$v\in\R^3$
and $P$ is a symmetric matrix such that $Pv=0$. Moreover,
$f$ is affinely conjugate to one of three possible normal forms,
depending on the dimension of $Z(v,L)$:
\begin{enumerate}
	 \item $\dim Z(v,L)=3$.
	 The map $f$ is conjugate to
	\begin{equation}\label{stdform}
	    \left({\begin{array}{c}\alpha+ \tau  x\ - \ \sigma y\ +\ z\ +\
Q(x,y)\\
	                            x\\
	                            y
	           \end{array}}\right)
	\end{equation}
	where $\tau$ and $\sigma$ are the trace and second trace of $L$, and
	$Q(x,y) = ax^{2}+bxy+cy^{2}$ is a quadratic form.

	  \item $\dim Z(v,L)=2$.
	The map $f$ is conjugate to
	 \[	\left({\begin{array}{c}x_0+ \alpha  x +y  +\ Q(x,z)\\
	                          y_0-  \beta  x\\
	                          z_0+  \frac1{\beta}z
	           \end{array}}\right).
	\]

	 \item $\dim Z(v,L)=1$.
	The map $f$ is conjugate to
	\[	\left({\begin{array}{c} x_0+\alpha  x +\ Q(y,z)\\
	                            y_0-\frac1{\alpha}z\\
	                            z_0+y+\beta z
	           \end{array}}\right).
	\]
\end{enumerate}
\end{teo}

\proof

We know that $f = L(x+\frac12 (x^{T}Px)x)+b$, and $Pv=0$. To obtain
the first normal form, perform a linear change of coordinates, $x=U
\xi$.  Since the vectors ${v}, {Lv},$ and ${L}^{2}{v}$ are linearly
independent, the transformation ${U}$ can be defined by the following
equations
\begin{xalignat*}{2}
	U^{-1}v &= {e}_{3} & \qquad {U}{e}_{3}& = v \\
	U^{-1}Lv&= {e}_{1} & \qquad {U}{e}_{1}& = Lv\\
	U^{-1}{L}^{2}v &= {e}_{2}  + \ \tau { e}_{ 1}& \qquad
	{U}{e}_{2}& = {L}^{2}v -  \tau  Lv
\end{xalignat*}
where, as we will see below, we will choose
$  \tau = Tr({L})$. In the new coordinates the map becomes
\begin{align*}
	\xi ' =& U^{- 1}f({U}( \xi  ))\\
	      =& U^{- 1}b + U^{- 1}L{U} \xi
	      +\frac12\left(\xi^TU^TPU\xi\right)U^{-1}Lv \\
	      =& { \xi }_{o} +U^{-1}L{U} \xi + {e}_{1}\tilde{Q}( \xi  , \xi  )
\end{align*}
where $\tilde{Q}( \xi_1 , \xi_2)=\frac12\left(\xi^T_1U^TPU\xi_2\right)$.
Note that $\tilde{Q}( \xi,e_3)=\frac12\left(\xi^TU^TPv\right)=0$,
 so in the new coordinates the quadratic terms depend only on the
first and second components. Moreover in this coordinate system
\begin{align*}
	U^{-1}L{U}{e}_{1}  =& U^{-1}{L}^{2}v = e_2 +  \tau e_1\\
	U^{-1}L{U}{e}_{2}  =& U^{-1}\left({{L}^{3}v- \tau { L}^{ 2} v}\right)\\
	U^{-1}L{U}{e}_{3}  =& U^{-1}Lv = {e}_{1}
\end{align*}
The second equation can be simplified by noting that the
characteristic equation for the matrix ${L}$ is satisfied by ${L}$
itself, and so ${L}^{3}- \tau {L}^{2}+ \sigma {L} - I =$ 0, where $
\tau = Tr({L})$ and $\sigma= Tr_{2}({L})$, the ``second trace'' of the
matrix $L$, thus we get $U^{-1}LUe_2= {U}^{-1}(I- \sigma{L})v =e_3-
\sigma e_1$. Thus we obtain
\[U^{-1}L{U} = \left({\begin{array}{ccc}
                                    \tau &-\sigma &1\\
                                     1   &0       &0\\
                                     0   &1       &0
                         \end{array}
			}
		   \right)\]
Upon reverting to $({x,y,z})$  as the names for the
coordinates we get
\[
U^{-1}f(U(x)) =
   \left({\begin{array}{c}
                  x_0\\
                  y_0\\
                  z_0
   \end{array}}\right) +
    \left({\begin{array}{c} \tau  x\ - \ \sigma y\ +\ z\ +\ Q(x,y)\\
                            x\\
                            y
           \end{array}}\right)
\]
To simplify this map further, we can conjugate, using the translation
\[
   (x,y,z)\mapsto (x,y+y_0,z+y_0+z_0),
\]
to a map with $x_0=\alpha$, $y_0=0$ and $z_0=0$. This is the promised
form.

For the second case, assume that $L^2v=\alpha Lv-\beta v$, for some 
nonzero $\alpha$ and $\beta$.  This implies that the characteristic
polynomial for $L$ factors as $(L-1/\beta I)(L^{2}-\alpha L +\beta I) = 0$,
and
therefore, since $L$ is nondegenerate, there exists a vector $w\notin
Z(v,L)$ such that $L w=\frac{1}{\beta}w$.  We define the following
change of coordinates.
\begin{xalignat*}{2}
	U^{-1}v &= {e}_{2} & \qquad {U}{e}_{2}& = v \\
	U^{-1}Lv&= {e}_{1} & \qquad {U}{e}_{1}& = Lv\\
	U^{-1}w &= {e}_{3} & \qquad {U}{e}_{3}& = w
\end{xalignat*}
As before, we note that in the new coordinates the quadratic term
satisfies $\tilde{Q}(e_2, \xi )=0$, so in the new coordinates the
quadratic terms depend only on the first and third components.
Moreover in this coordinate system we obtain
\[
    U^{-1}L{U} = \left({\begin{array}{ccc}
                                      \alpha &1  &0\\
                                    -\beta   &0  &0\\
                                         0   &0  &\frac1\beta
                         \end{array}
		    	}\right)
\]
This implies the form for the second case.

For the third case, assume that $Lv=\alpha v$.  We notice that there
exist a vector $w\notin Z(v,L)$ such that $Z(w,L)\oplus Z(v,L)=\R^3$.
In fact, we can also find a constant $\beta$ such that $L^2w-\beta
Lw+\frac1\alpha w=0$.  We define the following change of coordinates.
\begin{xalignat*}{2}
	U^{-1}v &= {e}_{1} & \qquad {U}{e}_{1}& = v \\
	U^{-1}w &= {e}_{2} & \qquad {U}{e}_{2}& = w\\
	U^{-1}Lw &= {e}_{3} & \qquad {U}{e}_{3}& = L w
\end{xalignat*}

As before, we note that in the new coordinates the quadratic term is
$\tilde{Q}(e_1, \xi )=0$, so in the new coordinates the quadratic
terms depend only on the second and third components.  Moreover in
this coordinate system we obtain
\[
     U^{-1}L{U} = \left({\begin{array}{ccc}
                                     \alpha &0  &0\\
                                        0   &0  &-\frac1\alpha\\
                                        0   &1  &\beta
                         \end{array}
			      }\right)
\]

This implies the form for the last case.
\qed

\section{Dynamics}
The dynamics of the second and third cases of theorem \Th{normalform}
are essentially trivial.  In case ii), the $z$ dynamics decouples from
the $(x,y)$ dynamics.  There are four special cases:

\begin{itemize}
 \item $|\beta| \neq 1$.
 The plane $z=\beta z_{0}/(\beta-1)$ is invariant, and is either a
 global attractor ($|\beta| >1$) or repellor ($|\beta|<1$).
 On the plane the dynamics is linear.
 \item $\beta = 1$, $z_{0} \neq 0$.
 All orbits are unbounded.
 \item $\beta = 1$, $z_{0}=0$.
 Every plane $z=c$ is invariant, and the dynamics reduces to a two
 dimensional area preserving H\'enon map on each plane.
 \item $\beta= -1$.
 Each plane $z=c$ is fixed under $f^{2}$. Restricted to a plane,
 $f^{2}$ is the composition of two orientation reversing H\'enon maps.
\end{itemize}

For case iii) the $(y,z)$ dynamics is linear and decouples from the
$x$ dynamics.  Generically, there is an invariant line on which the
dynamics is affine. The invariant line can have any stability type.

\subsection{Generic case}
Equation \Eq{stdform} is the only nontrivial case.  In general this
map has six parameters, one from the shift, two from the linear
matrix (the two coefficients of its characteristic polynomial) and the
three coefficients of $Q$.  However, generically, two of these
parameters can be eliminated.

Write the quadratic form as $Q(x,y) = ax^{2}+bxy+cy^{2}$.
Generically $a+b+c \neq 0$ and we can we can apply a scaling
transformation to set $a+b+c =1$.  Similarly if $b+2c \neq 0$ the
parameter $\sigma$ can be eliminated using the diagonal translation
\[
   (x,y,z)\mapsto ( x+\gamma,y+\gamma,z+\gamma ) , \ \ \ \gamma = \sigma
/(b+2c)
\]
In this way, we obtain the final, generic form
\begin{equation}\label{generic}
	\left({\begin{array}{c}x'\\
			y'\\
			z'
	       \end{array}}\right)=
	    \left({\begin{array}{c}\alpha +\tau x+z+ax^{2}+bxy+cy^{2}\\
				x\\
				y\end{array}}\right) \ \ \ a+b+c=1
\end{equation}
There are four parameters in the system.  Even if $a+b+c=0$ and/or
$b+2c=0$, then other normalizations can be chosen to eliminate two of
the parameters in \Eq{stdform}.  We will not study these special
cases.

\subsection{Periodic orbits}

Generically we can assume that $a+b+c=1$ for
the quadratic form in \Eq{stdform}.
The map \Eq{stdform} has at most two fixed points
\begin{equation}\label{fixedpts}
x=y=z=x_{\pm} = 
\frac12 \left(-{\tau}+\sigma\pm \sqrt{(\tau-\sigma)^{2}-4\alpha}\right)
\end{equation}
born in a saddle node bifurcation at $(\tau-\sigma)^{2}-4\alpha=0$. The
characteristic polynomial of the linearized map at the fixed points is
\[
  \lambda^{3}-t \lambda^{2}+s \lambda -1 =0
\]
where the trace $t$ and second trace $s$ are
\begin{eqnarray*}
     t_{\pm} &=& \tau +(2a+b)x_{\pm} \\
     s_{\pm} &=& \sigma -(2c+b)x_{\pm} 
\end{eqnarray*}
We notice that 
\begin{eqnarray*}
     t_{\pm} -s_{\pm}&=& \pm \sqrt{(\tau-\sigma)^{2}-4\alpha} \\
     t_{\pm} -s_{\mp}&=& \pm (a-c)\sqrt{(\tau-\sigma)^{2}-4\alpha}. 
\end{eqnarray*}

\begin{figure}[ht]
    \begin{center}
    \epsfig{file=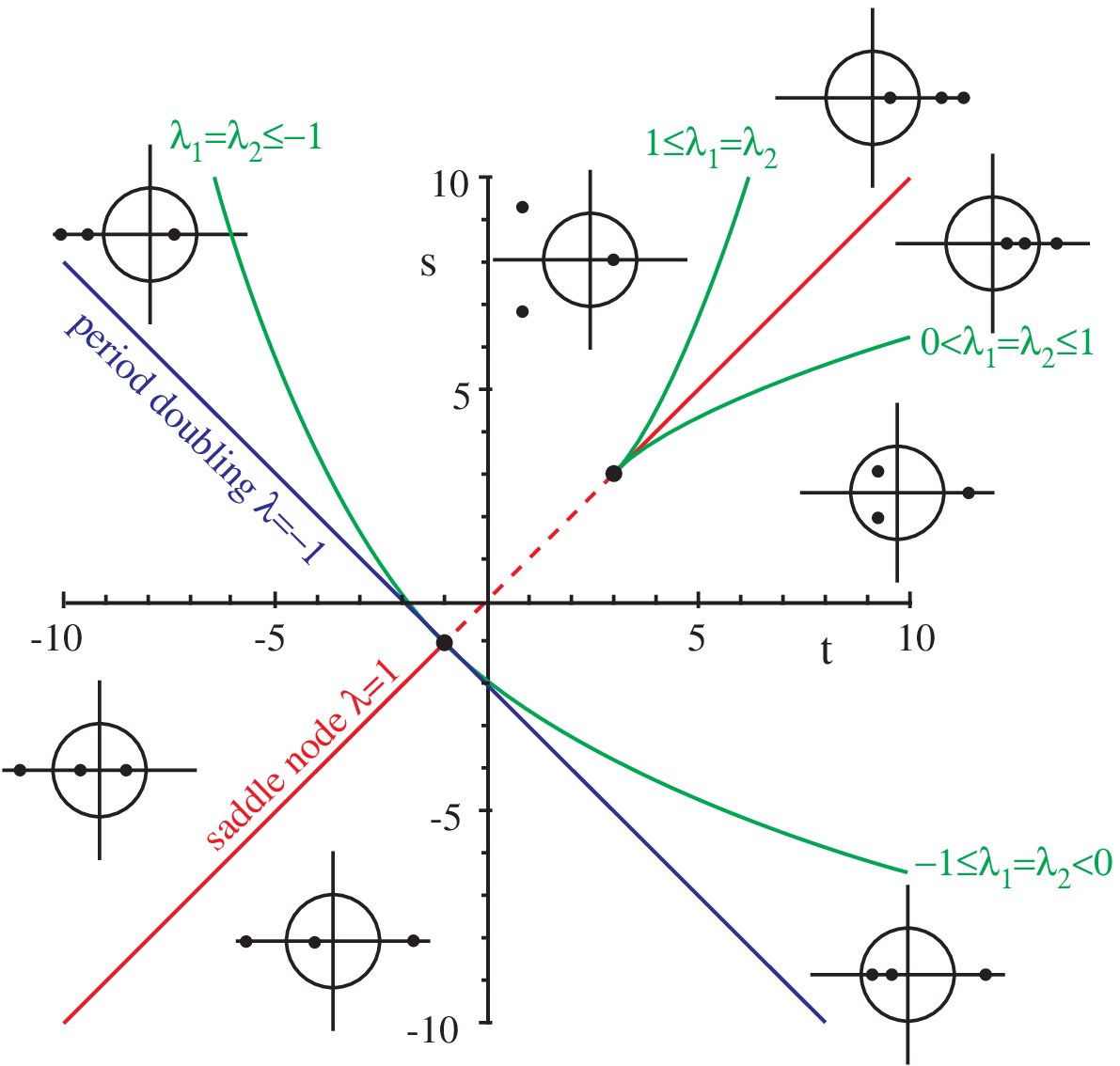,width=4.75in}
    \end{center}
    \caption{General stability diagram for a volume preserving map.}
     \label{ts-stab}
\end{figure}

The corresponding eigenvalues are illustrated in Fig \ref{ts-stab}.
It is easy to see (using the symmetric polynomials) that there are two
lines in $(t,s)$ space where the stability changes: the saddle-node
line $t=s$ corresponds to an eigenvalue $1$, and the period doubling
line $t+s=-2$ corresponds to an eigenvalue $-1$.  At the point
$t=s=-1$ where they cross the eigenvalues are necessarily $(-1,-1,1)$.
Note also that when $ -1 \leq t=s \leq 3$ there is a pair of
eigenvalues on the unit circle.  There are two other curves of
interest in the stability plane--these correspond to a double
eigenvalue $ \lambda_1 = \lambda_2 = r$, or
\[
	2r   +1/{r}^{2} = t \ \ \  r^{2}+ 2/r = s
\]
This gives the two curves shown in Fig \ref{ts-stab}. One has a cusp at
$t = s =3$, where we have the triple root $ \lambda=1.$ The second
crosses the saddle-node and period doubling lines at $t=s=-1$.  These
are the two codimension two points.

A fixed point with a one dimensional unstable manifold is called
{\em type A} and one with a one dimensional stable manifold is called
{\em type B}.  The saddle-node and period doubling lines divide the plane
into quadrants which alternate between type {\em A} and {\em B}.

Having a  pair of fixed points one of {\em type A} 
and one of {\em type B}, has interesting consequences for our map. For
instance, the  two dimensional manifolds serve as partial barriers to
transport.  Generically, they intersect  along a one dimensional manifold. We
have computed numerically some pairs of two dimensional stable and
unstable manifolds. As an example, see figure \ref{pescadito}.

We have noticed that varying the parameter makes the one dimensional
intersection bifurcate. Further investigation in this direction is
the subject of future papers and more complete treatment will appear
elsewhere.

The two fixed points \Eq{fixedpts} are born on the line
$t=s$ and move to opposite sides of this line for $(\tau-\sigma)^{2}>4\alpha$
($x_{+}$ is always on the right side).  If $|a-c|$ is small, they are
on the same side of the period doubling line, so that one is type {\em
A} and the other type {\em B}; however, when this parameter is
large enough they can be on opposite sides, and therefore of the same
{\em type}. This is determined by the sign of
\[
      s_{\pm}+t_{\pm}+2 = 2 + \tau+\sigma + 2(a-c)x_{\pm}
\]
When $a=c$, we have $t_{\pm}=s_{\mp}$ so that the eigenvalues of the
two fixed points are reciprocal (see section \ref{reverse} for the
explanation of this). 

\begin{figure}[ht] 
    \begin{center}
    \epsfig{file=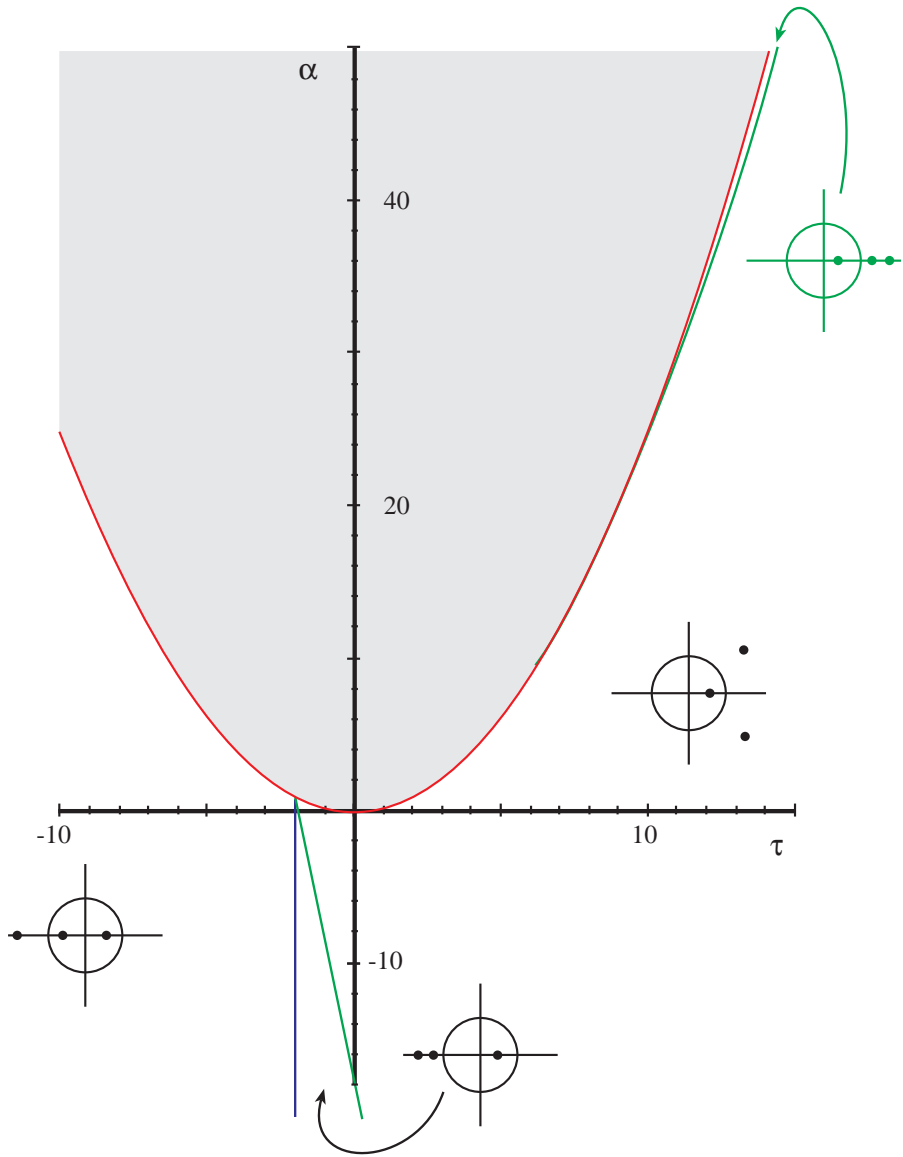,width=4.7in}
    \end{center}
    \caption{Stability Diagram for the reversible case $a+b+c=1$, $a=c$ 
    and $\sigma=0$.  There are no fixed points in the shaded region.  The 
    complex eigenvalues for the fixed point $x_{+}$ are shown.  Those of 
    $x_{-}$ are reciprocal to those of $x_{+}$.} \label{stab2}
\end{figure}

\begin{figure}[ht] 
    \begin{center}
    \epsfig{file=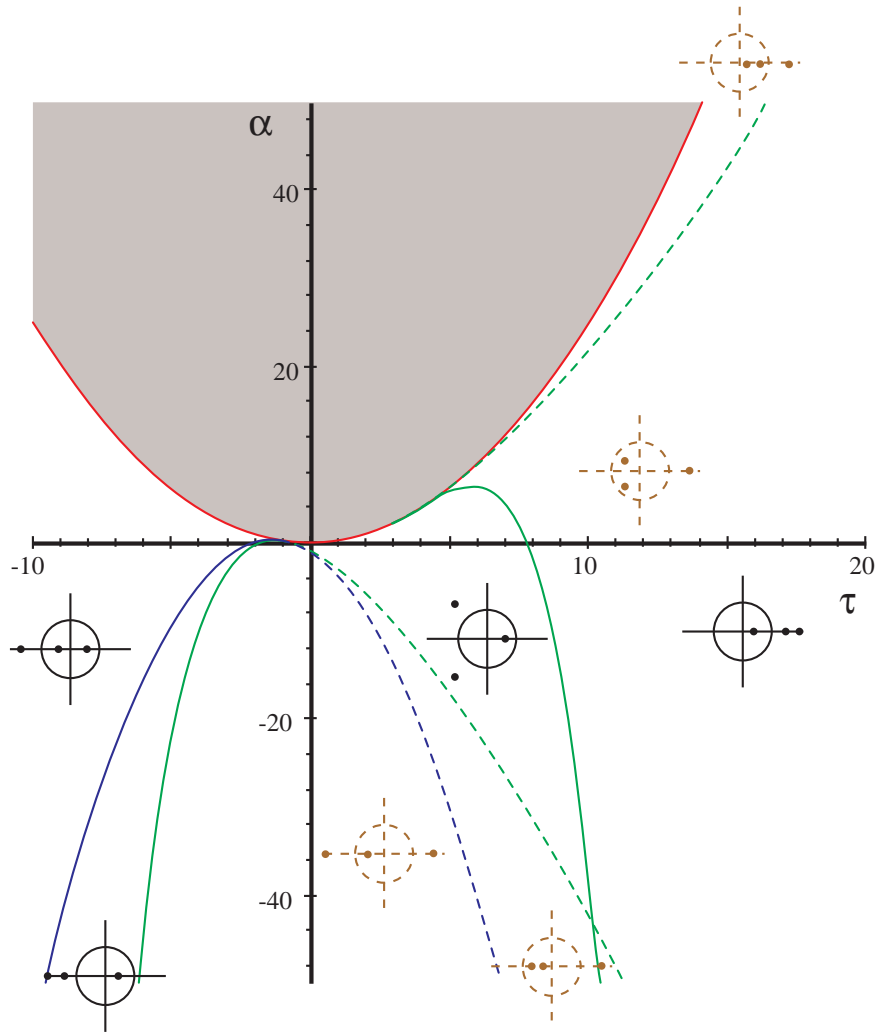,width=4.8in}
    \end{center}
    \caption{Stability Diagram for $ a=-1/2, b=1, c=1/2 ,\sigma=0$.
     Solid lines
    correspond to changes in the stability of $x_{+}$, while dashed
    lines to $x_{-}$.} \label{stab3}
\end{figure}

Remember that, generically and without loss of generality, we can assume
that $\sigma=0$. Therefore, we can plot stability diagrams for different 
values of $\tau$ and $\alpha$.
The stability diagram in the $(\tau,\alpha)$
plane for the $a=c$ case is shown in Fig \ref{stab2}. A more
general case is shown in Fig \ref{stab3}.

Periodic orbits can be studied by converting the map into a third
order difference equation.  Let $(x_{t},y_{t},z_{t}), t=0,1,\ldots$ be a
trajectory of the map \Eq{stdform}, then the map can be written as
\[
 x_{t+1}=\alpha+\tau x_{t}-\sigma x_{t-1}+x_{t-2}+Q(x_{t},x_{t-1}) \ .
\]
Now it is clear that if this system of $n$ quadratic equations has a
finite number of solutions, then there are at most $2^{n}$. We can
rule out this degeneracy is most cases

\begin{lem}
Suppose $a$ and $c$ are not both zero, and let $\mu_{\pm}$ be the two
(possibly complex) solutions of $Q(\mu,1)=0$. Then if
$\mu_{+}^{k}\mu_{-}^{n-k}
\neq 1$ for some integer $0 \leq k \leq n$, the number of fixed points of
$f^{n}$
for the map \Eq{stdform} is at most $2^{n}$.
\end{lem}

\proof
 
The ``nonlinear alternative'' \cite{Fried77} asserts that the
number of complex solutions, counted with multiplicity, of a system of
$n$ polynomial equations in $n$ variables is precisely the product of
the degrees of the polynomials providing the system of equations
obtained by discarding all terms but those of the highest degree in
each equation has only the trivial solution.  For our case, the
resulting system is
\[
     Q(x_{t},x_{t-1}) = ax_{t}^{2}+bx_{t}x_{t-1}+cx_{t-1}^{2} = 0,\
t=,1\ldots n, \ x_{n}=x_{0}  \ .
\]
If any one of the $x_{t}=0$, then they all are zero, unless $a=c=0$.
Otherwise the general nonzero solution to this system is
$x_{t}= \mu_{\pm} x_{t-1}$, where $a\mu^{2}+b\mu +c = 0$.
Setting $x_{n}=x_{0}$ requires $\mu_{+}^{k}\mu_{-}^{l}=1$, where $k+l=n$.
\qed

For example, typically there are at most four fixed points of $f^{2}$,
giving a single period two orbit in addition to the two fixed points
of $f$.  However, there could be infinitely many period two orbits or
none when $\mu_{+}\mu_{-}=1$, giving $a=c$ or $\mu_{\pm}^{2}=1$,
giving $b=\pm(a+c)$. As an example, when $a=c=b/2$ and
$\sigma+\tau+2=0$, the line $(x,\delta-x,x) \mapsto
(\delta-x,x,\delta-x)$ has period two where $\delta$ is defined by
\[
    \alpha + (1+\sigma)\delta + a\delta^{2} = 0
\]
\subsection{Reversibility} \label{reverse}

A map is reversible if it is conjugate to its inverse by a
diffeomorphism $h$ that is an involution, thus
\[
    h \circ f=f^{-1} \circ h  \ \ , \ \ h^{2}=I  \ .
\]
Some of the quadratic maps that we have considered have a reversor.
Assume the generic case $a+b+c\neq0$ (or equivalently $Q(1,1)\neq0$,
where $Q$ is the quadratic form given in \Eq{stdform}).
It is easy to see that,
if $a=c$,  then the map \Eq{stdform} has a reversor given by
\[
   h(x,y,z)=-\left({\begin{array}{c} z+ \eta \\
                                     y+ \eta \\
                                     x+ \eta
              \end{array}}\right) .
\]
where $\eta={  \tau -\sigma }/{(a+b+c)}.$

Note that when $f$ is reversible and has fixed points,
then the two fixed points have reciprocal
eigenvalues--so if one is type A, the other is type B. Moreover, if the
eigenvalues are complex, then the rotation rates have the same
magnitudes at the two fixed points.


\begin{lem} Let $f$ be a quadratic map in normal form \Eq{stdform}.
Assume, generically,  that the quadratic form satisfies $Q(1,1)\neq0$ and
$(\tau-\sigma)^2\neq 4\alpha\, Q(1,1)$
 Then $f$  is smoothly reversible if and only if $Q(x,y)=Q(y,x)$.
\end{lem}

\proof

Without loss of generality, we assume that $Q(1,1)=a+b+c=1$ and
$(\tau-\sigma)^2-4\alpha\neq0$.
Extend the map to $\C^3$. The imposed conditions imply that
the map $f$ has exactly two fixed points in $\C^3$.
 Suppose $f$ is reversible and has a fixed points $x_{\pm}$, then
it is easy to see that $h(x_{\pm})$ are also  fixed points.  
In addition, $Df(x_{\pm})$ is conjugate to $Df^{-1}(h(x_{\pm}))$.

Since there are two fixed points,  either $h(x_{\pm})=x_{\pm}$ or
$h(x_{\pm})=x_{\mp}$.  Now the eigenvalues are invariant under a
diffeomorphism, so in the first case the eigenvalues of $Df(x_{+})$
must be the same as those of $Df^{-1}(x_{+})$, and similarly for
$x_{-}$.   This can only happen when $t_{\pm}=s_{\pm}$
but this implies $x_{+}=x_{-}$, which can not happen by assumption.  

We conclude that
$h(x_{+})=x_{-}$.  This implies that $t_{\pm}=s_{\mp}$, which gives
$a=c$, or $Q(x,y)=Q(y,x)$.

The other direction is proved by a simple computation, as described above.
\qed

Reversibility simplifies finding orbits of a map. Orbits that are
invariant under $h$ are called symmetric, and as is easy to see, they
must have points on the fixed set $Fix(h) = \{x\in\R^3: h(x)=x\}$. 
In our case
this is the line $x+z=-\eta, y=-\eta/2$, so a numerical algorithm for
finding for symmetric orbits involves a one dimensional search.

Similarly, if the stable manifold of one of the fixed points
intersects $Fix(h)$, then the intersection point is on a heteroclinic
orbit, for suppose $x \in Fix(h) \cap W^{s}(x_{+})$ , then $x \in
W^{u}(x_{+})$, because $h(f^{n}(x)) = f^{-n}(h(x)) = f^{-n}(x)$, so
\[
\lim\limits_{n \rightarrow \infty }{f}^{n}(x)={x}_{+}
    \ \ \Rightarrow \ \ \
        \lim\limits_{n \rightarrow \infty} h({f}^{n}(x))
        =\lim\limits_{n \rightarrow \infty }{f}^{-n}(x)
     = h({x}_{+})={x}_{-}
\]
Furthermore, suppose the
stable manifold is two dimensional, and has the normal vector
$\hat{n}$ at a point on $Fix(h)$, then $Dh(x)\hat{n}$ is the normal
to the unstable manifold at this point. This implies that the
curve of heteroclinic orbits is tangent to the direction $\hat{n} \times
Dh(x)\hat{n}$.

\subsection{Bounded orbits}

For the H\'enon map, it is well known that the set of bounded orbits
is contained in a square.  For the volume preserving case, we will
show that an analogue of this result also holds providing the quadratic
form $Q$ is positive definite:

\begin{teo} \label{cube}
If $Q$ is positive definite then there is a $\kappa > 0$ such that all
bounded orbits are contained in the cube $\{(x,y,z): |x| \le\kappa,
|y|\le\kappa,|z|\le\kappa\}$.  Moreover, points outside the cube go to
infinity along the +$x$ axis as $ t\rightarrow +\infty$ or the -$z$
axis as $t \rightarrow -\infty$
\end{teo}

\proof

 We start by writing the map in third difference form as
\[
 x_{t+1}=\alpha+\tau x_{t}-\sigma x_{t-1}+x_{t-2}+Q(x_{t},x_{t-1}) \ .
\]
Recall that a quadratic form $Q(x,y)=ax^{2}+bxy+cy^{2}$ is positive
definite {\it iff} $a>0, c>0$, and $ d \equiv ac - b^2/4 > 0$.  We
will use the bounds obtained from completing the square:
\begin{eqnarray*}
     Q(x,y) &=& \frac{d}{c} x^{2} + c(y+bx/2c)^2 \ge \frac{d}{c} x^2  \ , \\
            &=& (x+by/2a)^{2} +\frac{d}{a} y^2  \ge \frac{d}{a} y^2 \ .
\end{eqnarray*}
There are three cases to consider, depending upon the relative sizes of $x_t$,
$x_{t-1}$, and $x_{t-2}$:

\begin{itemize}
\item $|x_t| \ge max(|x_{t-1}|,|x_{t-2}|$).
The difference equation then gives
\begin{eqnarray*}
  x_{t+1} &\ge& Q(x_{t},x_{t-1})  - |\alpha|-|\tau x_{t}|-|\sigma x_{t-1}|-|x_{t-2}| \\
          &\ge&  \frac{d}{c} x_{t}^{2} -(|\tau|+|\sigma|+1)|x_{t}| - |\alpha| \ ,
\end{eqnarray*}
Now since $d/c> 0$ there is a constant $\kappa_{1}>0$, depending on
$\alpha,\tau,\sigma, a,b$, and $c$ such that when $|x_t|>\kappa_{1}$,
we have
\[
 \frac{d}{c} x_{t}^{2} - (|\tau|+|\sigma|+1)|x_{t}|-|\alpha| > |x_{t}| \ ,
\]
In this case, we have $x_{t+1}>|x_t|$. Noting that  we then have
$x_{t+1}>|x_{t}|\ge|x_{t-1}|$, we can recursively apply this
result to show that the sequence
\[
    x_{t+k} > x_{t+k-1} > \ldots >|x_{t}| > \kappa_{1}
\]
is monotone increasing. In fact, this sequence is unbounded; otherwise it
would have a limit $x_{t} \rightarrow x^{*} > \kappa_{1}$, and this point would have
to be a fixed point $x=y=z=x^{*}$ of the map. However, there are at
most two such points, $x_\pm$, and a simple calculation shows that
$\kappa_{1} > x_{\pm}$, so both fixed point points are excluded.
\item $|x_{t-2}| \ge max(|x_t|,|x_{t-1}|)$.
Inverting the difference equation and shifting $t$ by one yields
\[
x_{t-3} = x_t - \alpha - \tau x_{t-1} +\sigma x_{t-2} - Q(x_{t-1},x_{t-2}) \ .
\]
Thus we have
\begin{eqnarray*}
    x_{t-3} &\le& -\frac{d}{a} x_{t-2}^2 +|x_t|+|\alpha|+|\tau x_{t-1}|+|\sigma x_{t-2}| \\
            &\le& -\frac{d}{a} x_{t-2}^2 + (|\tau|+|\sigma|+1)|x_{t-2}|+|\alpha| \\\
            &<&  -|x_{t-2}| \ ,
\end{eqnarray*}
when $|x_{t-2}|>\kappa_{2}$, for a constant $\kappa_{2}$ chosen as
before, but with $d/c$ replacing $d/a$.  This implies that the
sequence $x_{t-k} < x_{t-k+1} < \ldots < -|x_{t-2}|$ is monotone decreasing, 
negative, and unbounded.
\item  $|x_{t-1}| \ge max(|x_t|,|x_{t-2}|)$.
In this case we will see that the orbit is unbounded in both directions of time. For
the forward direction, note that
\begin{eqnarray*}
  x_{t+1} &\ge& \frac{d}{a}x_{t-1}^2   -(|\tau|+|\sigma|+1)|  x_{t-1}| - |\alpha|\\
          &>&  |x_{t-1}| \ ,
\end{eqnarray*}
 when $|x_{t-1}| >\kappa_2$. Thus $x_{t+1} > |x_{t-1}| \ge |x_t|$, which
is the situation covered by (i), and we get a monotone
increasing sequence, providing $x_{t+1}>\kappa_{1}$.  Alternatively, note that
\begin{eqnarray*}
   x_{t-3} &\le& -\frac{d}{c} x_{t-1}^2 +(|\tau|+|\sigma|+1) |x_{t-1}| + |\alpha|\\
           &<& -|x_{t-1}| \ ,
\end{eqnarray*}
when $|x_{t-1}|>\kappa_{1}$.  This gives $x_{t-3}<
-|x_{t-1}|<-|x_{t-2}|$, so we are in the situation covered by (ii),
which implies that the sequence approaches $-\infty$  providing
$|x_{t-3}|> \kappa_{2}$
\end{itemize}

In conclusion, we have shown that an orbit is unbounded either as $t
\rightarrow \pm \infty$ providing it contains a point $x_{t}$ such
that $|x_{t}| > max(\kappa_{1},\kappa_{2}) \equiv \kappa$. Note that
$\kappa_{1}$ is a monotone decreasing function of $d/c$, so therefore
we can define $\kappa$ by using $d/max(a,c)$:
\[
\kappa = \frac{max(a,c)}{2d}\left( |\tau|+|\sigma|+2 +
\sqrt{(|\tau|+|\sigma|+2)^{2} + 4\frac{|\alpha|}{d} max(a,c)} \right)
\]

Finally, we investigate the asymptotic direction of an unbounded orbit. Recall
that $y_t = x_{t-1}$ and $z_t= x_{t-2}$.  Suppose that $|x_t|> |y_t|>
|z_t|> \kappa$, then each of the variables is eventually positive, so
the orbit moves to infinity in the positive octant.  Moreover, once
all components are positive, we have
\begin{eqnarray*}
        {\frac{x_{t+1}}{x_t}} &=&
        {\frac{Q(x_t,y_t)}{x_t}} + {\frac{\alpha +\tau x_t -\sigma y_t + z_t}{x_t}} \\
        &\geq& \frac{d}{c} x_{t} -(|\tau|+ |\sigma|+1)-{\frac{|\alpha|}{x_t}}\rightarrow \infty  \ .
\end{eqnarray*}
So the ratios $y_t/x_t = x_{t-1}/x_t$ and $z_t/y_t  = x_{t-2}/x_{t-1}$ go to
zero, and the orbit approaches the positive $x$-axis as $t \rightarrow \infty$.
Similarly if $|z_t|> |y_t|> |x_t|> \kappa $, then eventually all the
components are negative, and so the orbit moves to infinity in the
negative octant as $t \rightarrow -\infty$. Once all components are
negative, we have
\begin{eqnarray*}
\frac{z_{t-1}}{z_t}
          &=& \frac{x_{t-3}}{x_{t-2}}  \\
          &=& -\frac{Q(y_t,z_t)}{z_t}+{\frac{x_t-\alpha-\tau y_t+\sigma z_t}{z_t}} \\
          &\leq&  -\frac{d}{a}|z_t|+(|\tau|+|\sigma|+1)+\frac{|\alpha|}{|z_t|} \rightarrow -\infty  \ .
\end{eqnarray*}
This implies the orbit moves to $\infty$ along the negative $z$ axis.
\qed

\section{Conclusions}
We have studied a family of volume preserving maps with the property 
that all entries are quadratic polynomials. We showed that these conditions
imply that such maps are polynomial diffeomorphisms. 
Then we restricted ourselves
to quadratic maps whose inverse is also quadratic. The class of maps 
studied is related to an old conjecture about polynomial maps called the
Jacobian conjecture.

A definition of quadratic shears was introduced   and 
a characterization was given in general. 
A further characterization in three space
was applied  to find  a normal form for the family.
In three-space, the form of the generic case is similar in
form to the area preserving
H\'enon map and, generically, the map has two fixed points that can be
either {\em type A} or {\em type B}. 

In addition,
using our definition of quadratic shear and its characterization, we were
able to give a simpler proof of a theorem of Moser classifying quadratic
symplectic maps.

Once we have found the normal form for the quadratic volume preserving map,
there remain many enticing open problems. For example,  we plan further
computations to visualize the stable and unstable manifolds of the fixed
points. Often these manifolds intersect, 
enclosing a ball; however, this is not guaranteed. Moreover, the
heteroclinic intersections, which are generically curves, can fall in many
homotopically distinct classes. We suspect that there are bifurcations
between these classes, and that which occurs will depend, for example, on
the complex phase of the eigenvalue of the associated fixed point.
Heteroclinic orbits can be found most easily for the reversible case, as an
intersections should occur on the fixed set of the reversor

Another problem of interest is to obtain a characterization of quadratic
shears in higher dimensions similar to the one we obtained in three
dimensions. At this point, normal forms could be obtained using techniques
similar to the current paper.

Finally, as we discussed in the introduction, one of our main motivations
for characterizing the quadratic volume preserving maps is to study
transport. If the two fixed points have disparate types, and their two
dimensional manifolds intersect on a circle, then transport can be
localized to ``lobes'' similar to the two dimensional case \cite{MacKay}.
However, as Fig. \ref{pescadito} shows, the intersections can be curves that
spiral from one fixed point to the other.  We plan to characterize
transport for such cases. The existence, for the definite case,
of a cube containing the bounded
orbits (cf. theorem \ref{cube}) will prove useful in this study.

\clearpage
\bibliographystyle{unsrt}  
\bibliography{tev}

\clearpage
\section{Figures}

\begin{figure}[ht] 
    \vspace{4in}
    \caption{Two dimensional stable and unstable 
    manifolds for the parameters $ a=c=0.5, b=0.0, 
  \alpha=0.0,\tau=-0.3$ and $\sigma=0.0$.} \label{pescadito}
\end{figure}

\end{document}